\newcommand{\cappa}{ {\cal K} }   		% Schmidt number
\newcommand{\om}{\omega}			% Shift in frequenza
\newcommand{\Om}{\Omega}			% Shift in frequenza
\newcommand{\Oms}{\Omega_s}	
\newcommand{\Omi}{\Omega_i}

\newcommand{\DDpm}{{\cal D}^{\hspace{-0pt}\text{\tiny{$(l)$}}}}

\newcommand{\spj}{}
\newcommand{\spa}{}

\newcommand{\sppm}{}
\newcommand{\taus}{{\tau_{p,s}^{\text{\tiny{($-$)}}}}}
\newcommand{\taui}{{\tau_{p,i}^{\text{\tiny{($+$)}}}}}
\newcommand{\tausb}{{\tau_{p,s}^{\text{\tiny{($-$)}}}}}
\newcommand{\tauib}{{\tau_{p,i}^{\text{\tiny{($-$)}}}}}
\newcommand{\taujpm}{{\tau_{j}^{\text{\tiny{($\pm$)}}}}}
\newcommand{\tauipm}{{\tau_{p,i}^{\text{\tiny{($\pm$)}}}}}

\newcommand{\gamgauss}{\gamma}

\newcommand{\taup}{{\tau_\mathrm{p}}}
\newcommand{\as}{\hat{a}_s}
\newcommand{\ai}{\hat{a}_i}

\newcommand{\sinc}{{\rm sinc}}

\newcommand{\nn}{\nonumber}
\newcommand{\bsub}{\begin{subequations}}
\newcommand{\esub}{\end{subequations}}
\newcommand{\beq}{\begin{equation}}
\newcommand{\eeq}{\end{equation}}
\newcommand{\beqa}{\begin{eqnarray}}
\newcommand{\eeqa}{\end{eqnarray}}
\newcommand{\beql}{\begin{subequations}\begin{eqnarray}}
\newcommand{\eeql}{\end{eqnarray}\end{subequations}}
\documentclass[pra,aps,amsmath, twocolumn, nofootinbib,showpacs]{revtex4-1}
\usepackage{amsmath} 
\usepackage{empheq}
\usepackage{amssymb}
\usepackage{graphicx}
\usepackage{longtable}
\usepackage{float}   % per posizionare meglio le figure con [H]
\usepackage{verbatim}  %multi-line comment \begin{comment}....\end{comment}
\usepackage{braket}
\usepackage{color}
\begin{document}
\title{Heralding pure single photons: a comparision between counter-propagating and co-propagating twin photons.}
%Heralding pure single photons. A comparision between counter-propagating and co-propagating photon pairs
\author{E.~Brambilla$^2$, T. Corti$^2$ and A.~Gatti$^{1,2}$ }
\affiliation{$^1$ Istituto di Fotonica e Nanotecnologie del CNR, Piazza Leonardo  Da Vinci 32, Milano, Italy;  \\
$^2$ Dipartimento di Scienza e Alta Tecnologia dell' Universit\`a dell'Insubria, Via Valleggio 11,  Como, Italy}
\begin{abstract}
We investigate the possibility of generating pure heralded single photons through spontaneous parametric down-conversion comparing the counter-propagating geometry studied in \cite{gatti2015} with more conventional co-propagating configurations which enhance the purity of the heralded photon state through the technique of group-velocity matching. 
We estimate the Schmidt number associated to the temporal modes as a function of the pump pulse duration for three particular configurations, showing how the different phase-matching conditions influences the degree of separability that can be achieved.
\end{abstract}
%\centerline{Version \today}
\maketitle
%%%%%%%%%%%%%%%%%%%%%%%%%%%%%%%%%%555
\section*{Introduction}
\label{sec:intro}

In the process of spontaneous parametric down-conversion (SPDC) occurring in a $\chi^{(2)}$ material, 
photons belonging to the laser pump field are split
into pairs of photons of lower energies and momentum.
The generated photon pairs, are naturally entangled in a number of variables
(energy, momentum, angular momentum, polarization)
as a consequence of the conservation laws ruling the microscopic  process.
Because of its relative simplicity of implementation, SPDC is indeed a widely used  source of entangled light.
%and has been employed in a number of applications based on entanglement, such as e.g. quantum cryptography, quantum teleportation and dense coding.
At the same time, it is also the most frequently used source of pure photons heralded by the detection of their twin partner,
the starting point of many quantum information protocols.
In this latter case entanglement must be avoided as much as possible, 
since the heralded photons are required to be in indistinguishable and capable of high-visibility interference. 
Filtering is the simplest method for achieving this purpose, through it presents 
the drawback of drastically reducing the efficiency of the source
% since only a tiny fraction of the emitted twin photons are exploited. 
In order to achieve pure heralded photons with high fluxes, considerable effort has been devoted  
to find alternative techniques that do not rely on post-selection \cite{grice2001,uren2006,mosley2008b,migdall2002,levine2010,bennink2010,branczyk2011,zhang2012}. 
They consist in manipulating directly the degree of entanglement of the source by controlling
the modal structure of the emitted photon pairs in order to produce
uncorrelated twin photons 
%in a factorable state.
In such
a way a conditioned 
measurement projects the field in a pure single photon state rather than in a mixed state, and
pure heralded photon can be obtained without filtering. A recent survey of these techniques can be found in \cite{migdall2013}.
\par
In this work we investigate different configuration to eliminate entanglement in the temporal frequency domain, in particular comparing a co-propagating and a counter-propagating geometry. 
In the latter, proposed by Harris in the sixties \cite{harris66} and implemented in 2007 by Canalias et al. \cite{canalias2007}, twin photons
are emitted along opposite directions in a periodically poled crystal with a submicrometric poling period. 
The technological challenges involved in the fabrication of crystals with such a short poling period are described e.g. in \cite{canalias2003,canalias2012}.
With respect to the standard co-propagating configuration where the twin photons are typically emitted over a broad range of frequencies, 
counter-propagating photons have much narrower spectral bandwidths imposed
by the peculiar phase-matching conditions characterizing this geometry \cite{canalias2012b,suhara2010,gatti2015,corti2016}.
Because of this feature, the counter-propagating  SPDC configuration
has soon been recognized as a promising source for generating heralded single photons 
%characterized by a high level of purity and high fluxes at the same time
 \cite{christ2009,gatti2015}. 
A detailed analysis of the temporal coherence and correlation of counter-propagating twin photons and twin beams
has been performed in previous works of ours \cite{gatti2015,corti2016} , 
where we studied both the  spontaneous  regime \cite{gatti2015} and  the stimulated regime of photon pair
production \cite{corti2016}. 
%where we studied both the  purely spontaneous  regime \cite{gatti2015} and  the regime of stimulated production of %photon pairs where distributed feedback play a fundamental role \cite{corti2016}. 
\par
In this work we focus on the purely spontaneous regime, where the system can be exploited as a source of
heralded single photons, and we provide a detailed comparison between this source and the conventional  co-propagating configuration. In the latter case,  a separable two-photon state can be achieved only through the techniques of group velocity matching, which require a careful choice of the material and of the tuning conditions as well as sub-picosecond pump pulses \cite{grice2001}. 
Conversely, our analysis will show that in the counter-propagating geometry there is no need of such a fine tuning,  and that highly monochromatic single photons in a pure state can be generated in a wide range of  phase-matching conditions and pump durations. 
Conversely, our analysis will show that in the counterpropagating geometry there is no need of such a fine tuning,  and that highly monochromatic single photons in a pure state can be generated in a wide range of  phase-matching conditions and pump durations. 
In particular,  we shall emphasize how the different time scales in play characterizing the two configurations strongly affect the conditions in which separability can be achieved and are at the origin of the different behaviors of the two sources.  
\par
The paper is organized as follows: Sec.\ref{sec:phasematching} illustrates the two geometries and their  different phase-matching conditions.  Sec.\ref{sec:Schmidt} 
evaluates the degree of entanglement of the  two-photon state,  providing approximated analytical expressions
for the Schmidt number, valid in both geometries. Examples of specific configurations suitable for generating pure heralded 
photons are analyzed in Sec.\ref{sec:examples} and compared with the counter-propagating geometry.  The spectral properties of co-propagating and counterpropagating twin photons are finally analysed in Sec.  \ref{sec:spectrum}. 

\section{Phase-matching in the counter- and co-propagating geometries}
\label{sec:phasematching}
We restrict our analysis to a purely temporal description:
we consider only collinear propagation, either assuming that
a small angular bandwidth is collected and the process is
characterized by a single spatial mode operation, or because
of a waveguiding configuration. 

\begin{figure}[h]
\includegraphics[scale=0.38]{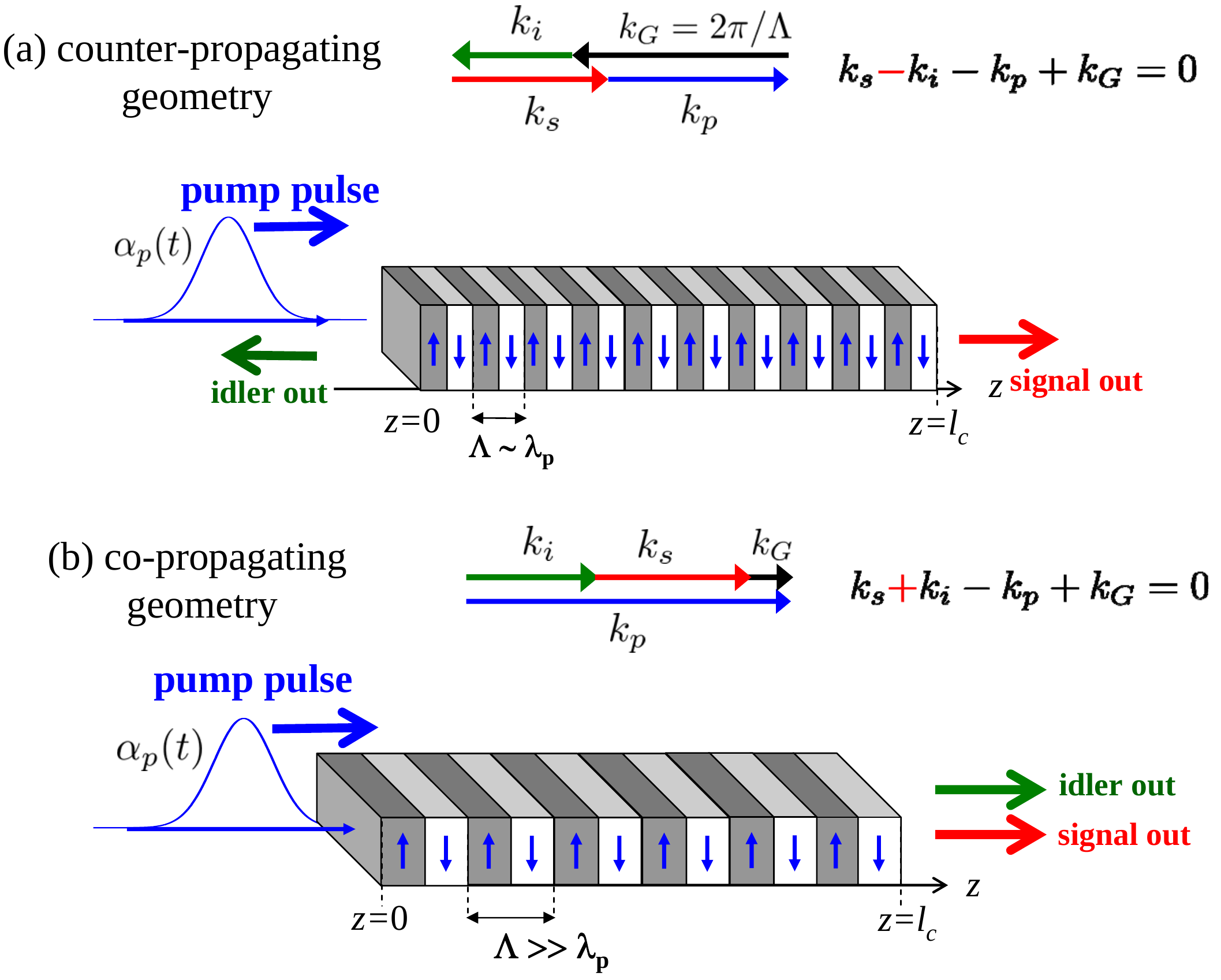}
%\caption{(a) Optical scheme for the generation of twin photons in the counter-propagating geometry and (b) 
%the corresponding quasi-phasematching condition for collinear propagation which requires a submicrometric
%poling period such that $k_G\sim k_p$. For comparison, (c) illustrates
%the quasi-phasematching conditions for the usual co-propagating geometry in which the poling period $\Lambda$ is 
%typically in the 10-100 micrometer range.}
\caption{(a) Scheme of  twin-photon  generation  in the (a)  counter-propagating and (b) 
co-propagating geometries. 
In case (a) the quasi-phasematching for collinear propagation requires a submicrometric
poling period,  $k_G\sim k_p$, while the co-propagating  process (b) can be phase matched either in bulk crystals,  or with larger poling periods $\Lambda$ .}
\label{fig1}
\end{figure}

We first consider the counter-propagating geometry shown in Fig.\ref{fig1}a, with 
a coherent pump pulse of central frequency $\omega_p$ and temporal profile $\alpha_p(t)$ impinging a periodically poled crystal 
of length $l_c$ from the left face
and generating counter-propagating photon pairs with, say,  the idler photon  
propagating opposite to the pump.   
This occurs when the poling period $\Lambda$  is on the same order of the pump wavelength 
in the medium $\lambda_p/n_p$: in this case 
the  first-order momentum associated to the nonlinear grating, $k_G=2\pi/\Lambda$ 
is able to compensate  the pump photon momentum so that twin photons must be emitted
along opposite directions in order to satisfy momentum conservation (Fig.\ref{fig1}a). The central frequencies of the emitted signal
and idler fields, $\omega_s$ and $\omega_i=\omega_p-\omega_s$, 
are thus determined by the crystal poling period $\Lambda$ and the pump central frequency $\omega_p$
according to the following quasi-phasematching condition
\beq
\label{qpm}
k_{s}-k_{i}=k_{p}-k_G\;\;\text{\small{{\bf (a) counter-propagating case}}}
\eeq
where $k_{j}:=\frac{\om_j}{c}n_j(\omega_j)$, $j=s,i,p$ denotes the wave-number
at the corresponding central frequencies $\omega_j$.

For comparison, we shall also  consider the more common co-propagating geometry (Fig.\ref{fig1}b) 
where the all the three   fields propagate  along the positive $z$ direction. In this case the wave-numbers
at the reference frequencies satisfy the following  condition
\beq
\label{qpm2}
k_{s}+k_{i}=k_{p}-k_G\;\;\text{\small{{\bf (b) co-propagating case}}}
\eeq
which  can describe   both  the case of a bulk crystal, in which   $k_G=0$,  or quasi-phasematching in periodically poled
structures with $k_G\ll k_j$, $j=i,s,p$. 

Considering the regime of photon-pair production, 
the generated two photon state conditioned by a photon count has the form \cite{gatti2015}
\beq 
|\phi^{\spj}_{\mathrm C}\rangle =  \int d\Om_s d\Om_i \psi^{\sppm} (\Om_s, \Om_i) \as^\dagger (\Om_s) \ai^\dagger (\Om_i) \left |0\right \rangle 
\label{statec}
\eeq
where
$\as^\dagger (\Om_s)$ and $\ai^\dagger (\Om_i)$ denote  the signal and idler creation operators in the frequency domain ($\Om_j$ is the offset from the reference frequency $\omega_j$), and
\beqa
\label{psi}
&&\psi^{\sppm}(\Omega_s,\Omega_i)
=\frac{g}{\sqrt{2\pi}}
\tilde{\alpha}_p(\Omega_s+\Omega_i)\nn\\
&&\times e^{-i\frac{1}{2}\DDpm(\Omega_s,\Omega_i)l_c}
{\rm sinc}\left[\frac{\DDpm(\Omega_s,\Omega_i)l_c}{2}\right],\;\;(i=a,b)
\eeqa
is the so-called biphoton amplitude, giving the joint probability amplitude of detecting a signal photon at frequency $\omega_s+\Om_s$
and an idler photon at frequency $\omega_i+\Om_i$. Here, 
$\DDpm(\Oms,\Omi)l_c$ denote the phase-mismatch functions, where the $i=a, b$ superscripts refer to the counter-propagating and the co-propagating 
configuration respectively. 
%\beqa
%a_s(\Oms)=\As(\Oms)
%\eeqa
$\tilde{\alpha}_p(\Om)$ is the  spectral amplitude of the pump pulse normalized by its peak value
\beq
\label{alphap}
\tilde{\alpha}_p(\Om)=\int\frac{dt}{\sqrt{2\pi}}e^{i\Om t}\frac{\alpha_p(t)}{\alpha_p(t=0)}\,,
\eeq
The differences between the two geometries arise because of   the different sign characterizing 
the two phase-mismatch functions 
\beqa
&&\DDpm(\Oms,\Omi) \nonumber \\
&&=
\left\{
\begin{array}{lr}
k_s(\Oms)-k_i(\Oms)-k_p(\Oms+\Omi)+k_G,&\text{ $l=a$}\\
k_s(\Oms)+k_i(\Oms)-k_p(\Oms+\Omi)+k_G,&\text{ $l=b$}
\end{array}
\right.
\label{mismatch}
\eeqa
%Here and in the following we shall use the superscript {\tiny{$^{(-)}$}} for the quantities involved in the co-propagating
%geometry where $\tauib$ enters into play, the superscript {\tiny{$^{(+)}$}} for corresponding quantities in the counter-propagating %geometry which depends on $\taui$.

As extensively  described in \cite{gatti2015, corti2016}, the properties of the counter-propagating twin photons 
strongly differ from those of the  co-propagating ones, because of the minus sign in front of the idler wave-number $k_i(\Om_i)$. 
This is best seen by expanding the phase mismatch (\ref{mismatch}) at first order around the reference frequencies (corresponding to $\Omega_j=0$), obtaining thereby
\beqa
\label{Dlin}
\frac{1}{2}\DDpm(\Oms,\Omi)l_c
&\approx&\frac{l_c}{2}[(k_s'- k_p')\Oms-(k_p'\mp k_i')\Omi]\label{Dlin1}\\
&\equiv& -\left(\taus\Oms+\tauipm\Omi\right)\;,\label{Dlin2}
\eeqa
where the $+$ and $-$ minus sign refer to the counter-propagating  $(l=a)$ and co-propagating  $(l=b)$ cases, respectively,
$k_j'\equiv v_{gj}^{-1} =\left(\frac{d k_j}{d \Omega_j}\right)_{\Omega_j=0}$
is the inverse group velocity of wave $j$ at its reference frequency $\omega_j$ and the  characteristic times
\beqa
\label{taujpm}
\taujpm
=\frac{1}{2}\left(\frac{l_c}{v_{gp}}\pm\frac{l_c}{v_{gj}}\right)\;\;\;j=i,s
\eeqa
involve either the difference or the sum of the inverse group velocities  
of the pump  and of the wave $j=i,s$, depending whether the latter copropagates or counterpropagates with respect to the former. 
The omission of higher order dispersion terms is  justified in the counter-propagating configuration,  which typically involves narrow down-conversion spectra \cite{canalias2007,gatti2015, corti2016}. In order to perform analytical calculations, we shall use the linear approximation \eqref{Dlin2} also for the co-propagating configuration, through in this case the effect of group velocity dispersion can be more relevant because of the larger bandwidths in play.

As discussed in \cite{grice2001,uren2006,grice2010,gatti2015}, the possibility to generate
heralded photons with a high degree of purity depends  both on the relative sizes and signs of the time constants $\taus$and $\tauipm$ defined in Eq.\eqref{taujpm},  and on how they compare
to the pump duration $\tau_p$
(notice that only $\taui$  is always positive, while $\taus$ and $\tauib$ can be either positive or negative).  
In the co-propagating geometry the two  scales $\taus$ and $\tauib$ are associated to the group velocity mismatch (GVM) of the signal and the idler  with respect to the co-propagating pump. They are on the same order of magnitude,  except for the particular case in which one of the two fields is velocity matched to the pump.
By contrast, in the counter-propagating case, the time constant associated to the  backward photon
$\taui=l_c/2v_{gp}+l_c/2v_{gi}$, which involves  
inverse group velocities sum (GVS), is  on the order of the photon transit time across the crystal, and 
exceeds therefore the signal-pump GVM  time $\taus=l_c/2v_{gp}-l_c/2v_{gs}$ by at least an order of magnitude. 

Therefore, considering the ratio between the two time constants , 
\begin{equation}
\eta^{\spa}=\taus/\tauipm \;,
\end{equation}
we have 
\beq
|\eta|= \begin{cases} 
|\taus|/|\taui|\ll 1\;  &\text{for case (a)} \\
|\tausb|/|\tauib| \; &\text{arbitrary for case (b)}.
\end{cases}
\label{eta2}
\eeq
Without loss of generality, in case (b) 
%we shall assume the pump group velocity is closer to the
%signal group velocity than to the idler group velocity, i.e. $|\tausb|<|\tauib|$, 
we shall assume that the signal and idler fields satisfy the condition $|\tausb|\leq|\tauib|$
so that in both configurations we have
\beq
-1\leq\eta^{\sppm}\leq 1
\eeq

\section{Entanglement quantification}
\label{sec:Schmidt}
We characterize the degree of entanglement with the Schmidt number \cite{ekert1995,parker2000}, which provides an estimate of the number modes participating 
to the entangled state \cite{exter2006}. 
It is defined as the inverse of the purity of the state of each separate subsystem
\beq
\cappa= \frac{1}{ {\mathrm Tr} \{\rho_s^2\} }= \frac{1}{ {\mathrm Tr} \{\rho_i^2\} }
\eeq
where $\rho_s$,  $\rho_i$  are the reduced density matrix of the signal  and  idler , e.g. $\rho_s= {\mathrm Tr}_i \{  |\phi _{\mathrm C}\rangle \, \langle 
\phi _{\mathrm C}|  \} $. 
For a biphoton state of the form \eqref{statec}, 
the Schmidt number can be expressed through integrals involving the first-order
coherence functions of the signal and the idler fields
\beqa
&&G_s^{(1)}(\Oms,\Oms')=\int d\Omi \psi^*(\Oms,\Omi)\psi(\Oms',\Omi) \label{G1s}\\
&&G_i^{(1)}(\Omi,\Omi')=\int d\Omi \psi^*(\Oms,\Omi)\psi(\Oms,\Omi') \label{G2s}
\eeqa
Namely, it is found that \cite{gatti2012,mikhailova2008}
\beq
\cappa = \frac{ {\cal N}^2} {B}   
%\frac{ \langle \hat{N_j} \rangle ^2} { \langle: \delta \hat{N}_j^2: \rangle }    \qquad j=s,i
\label{kintegral}
\eeq
where
\begin{align}
 {\cal N}  = & 
\int d\Om   \, G_s^{(1) }(\Om, \Om)     = \int d\Om   \, G_i^{(1) }(\Om, \Om) 
%= \int d\Om_i  \left|G_i(\Om_i, \Om_i)\right|^2  ,
\label{enne} \\
B= & \int d\Om \int  d\Om' \left|G_s^{(1) }(\Om, \Om')\right|^2 \nonumber  \\
= & \int d\Om \int  d\Om' \left|G_i^{(1) }(\Om, \Om')\right|^2     
\label{B} 
%= &\int d\Om_s ... \int d\Om_i'   \left[  \psi ( \Om_s,\Om_i)  \right. 
 %                     \psi (\Om_s',\Om_i')    \psi^* (\Om_s,\Om_i')     \nn    \\
%&     \qquad \qquad  \qquad \qquad \qquad \qquad \times\left.  \psi^* (\Om_s',\Om_i)  \right]  
%\label{NB}
\end{align}

%\section{Estimation of the Schmidt number}
%\label{sec:Kgauss}
In this work the Schmidt number $\cappa$ will be estimated  by means of  i) the numerical integration 
of  Eqs.(\ref{kintegral})-(\ref{B}),  where  the phase-mismatch  (\ref{mismatch}) is evaluated using the complete Sellmeier dispersion formula in \cite{niko91,kato2002,zernike64}, or ii)  a  Gaussian approximation for the biphoton amplitude  in Eq.(\ref{psi}),  which is typically used in the literature \cite{grice2001,law2004,uren2006}.
In the latter case,   the sinc function in Eq.(\ref{psi}) is fitted by  a Gaussian of its argument, setting
\begin{align}
\sinc\frac{\DDpm(\Oms,\Omi)l_c}{2}&\approx e^{-\gamgauss\left(\frac{\DDpm(\Oms,\Omi)l_c}{2}\right)^2}\\
&\approx  e^{ -\gamma \left(\taus\Oms+\tauipm\Omi\right)^2}
\label{sinc_apr}
\end{align}
where  the linear approximation \eqref{Dlin}   of the phase mismatch has been used  in the second line.  
$\gamgauss$ is a  fitting parameter; e.g. 
requiring  that the sinc and the Gaussian functions shares the same full width at half maximum (FWHM), one has  $\gamma=0.193$.  The approximation \eqref{sinc_apr} allows to derive analytical results,  which provide  an immediate comparison between the various configurations, but  neglects the effects  of group velocity dispersion at second and higher  orders. 

Furthermore, we consider a pump pulse with a Gaussian temporal profile  of duration $\tau_p$, 
$\alpha_p(t)=\alpha_p(0)e^{-t^2/2\tau_p^2}$ , so that the corresponding 
spectral amplitude \eqref{alphap} is given by
\beq
\tilde{\alpha}_p(\Om)=\frac{1}{\Delta \Om_p} e^{-\frac{\Om_p^2}{2\Delta \Om_p^2}},
\eeq
 where the spectral bandwidth   is  $\Delta \Om_p=1/\tau_p$.\\
The  biphoton amplitude (\ref{psi})  takes then the approximated form : 
\beqa
\label{psi_gauss}
\psi^{\sppm}(\Oms,\Omi)\approx &&
\frac{g\tau_p}{\sqrt{2\pi}}
e^{i\left[\taus\Oms+\tauipm\Omi\right]}\nn\\
&&\times e^{-c_{11}\Oms^2-c_{22}\Omi^2-2c_{12}\Oms\Omi} 
\eeqa
where the real coefficients $c_{ij}$  are
\beqa
c_{11}&=&\frac{\tau_p^2}{2}+\gamgauss\taus^2,\label{c11}\\
c_{22}&=&\frac{\tau_p^2}{2}+\gamgauss\tauipm^2,\label{c22}\\
c_{12}&=&\frac{\tau_p^2}{2}+\gamgauss\taus\tauipm.\label{c12}
\eeqa
Here and in the following $\taui$ refers to the counter-propagating case ($a$), $\tauib$ refers to 
co-propagating case ($b$). 
Inserting  approximation \eqref{psi_gauss} in the expression  of  $\cappa$ in  Eqs.\eqref{kintegral}-\eqref{B}, we find
\begin{align}
\label{Kgauss}
\cappa^{\sppm} &=\sqrt{\frac{c_{11}c_{22}}{c_{11}c_{22}-c_{12}^2}}\\
&=
\frac{1}{1-\eta^{\sppm}}
\left[
1+\eta^2
+\frac{1}{2\gamgauss}\left(\frac{\tau_p}{\tauipm}\right)^2
+2\gamgauss\left(\frac{\taus}{\tau_p}\right)^2
\right]^{1/2}
\end{align}
As a function of the pump duration $\tau_p$, it is easily seen that $\cappa^{\sppm}$ takes its minimum for
\beq
\label{taupmin}
\tau_{p}^{min} 
=\sqrt{2\gamgauss|\taus\tauipm|}
\eeq
The minimum value of $\cappa$ depends both on the sign and on the magnitude of $\eta^{\sppm}$ and is given by
\beq
\cappa_{min}^{\sppm}
=
\left\{
\begin{array}{ll}
\frac{1+\eta^{\sppm}}{1-\eta^{\sppm}}&\text{for $\eta^{\sppm}>0\rightarrow \taus\tauipm > 0$}\\
1&\text{for $\eta^{\sppm} \leq 0\rightarrow \taus\tauipm \leq 0$}
\end{array}
\right.
\label{Kmin}
\eeq
Thus, within the validity of the Gaussian approximation (\ref{psi_gauss}), 
complete separability can be achieved only if $\eta \leq 0$. 
The condition $\eta=0$ corresponds to perfect velocity matching between the pump and the signal:   $\taus=0$.  Notice that in this case perfect separability is reached only asymptotically for  $\tau_p \to \tau_{p}^{min} =0$. 
Conversely,  when  $\eta <0$,  perfect separability $\cappa=1$ can be in principle  reached  for finite pump durations, and requires 
that  $\taus$ and $\tauipm$ have opposite signs. 
Once this condition is met,  the mixed term coefficient
(\ref{c12}) vanishes for a pump duration  $\tau_p= \tau_p^{min}$. \\
Alternatively, for positive $\eta$, the two-photon state can be made almost separable by choosing a
configuration for which $\eta$ is sufficiently small.  Notice that this last condition is naturally fulfilled in the counter-propagating case. 
\section{Specific configurations for separability } 
\label{sec:examples}
According to the results presented in Sec.\ref{sec:Schmidt}, we shall compare three distinct configurations which satisfy the  conditions for complete or nearly complete separability: 

{\bf (i) Counter-propagating geometry ($|\eta|<<1$)}\\
The peculiarity of the counter-propagating geometry (a) is  that the condition $|\eta^{\spa}|\ll 1$ is naturally fulfilled 
[see Eq.(\ref{eta2}) and discussion].
Even for $\eta^{\spa} >0 $ , an almost separable state can be always reached, because the minimum of 
$\cappa^{\spa}$  is, 
\beq 
\cappa_{min}^{\spa}\approx 1+2\eta^{\spa}\approx 1
\eeq 
and $\cappa$ stays close to this value within
a broad range of pump durations  around   $\tau_p^{min}$, which 
is basically  the geometric mean of $\taus$ and $\taui$ (see  Eq.(\ref{taupmin})). 
As already noticed, for a few millimeter crystal  
$\taui$ is on the order of  several tens of picoseconds while $\taus$ and $\tauib$ 
are typically in the subpicosecond range. 
The required pump duration $\tau_p^{min}$ is thus on the order of several picoseconds, and  thus is easily accessible and 
significantly longer than in the co-propagating configurations considered next. 

These results are in agreement with the more general analysis presented in  \cite{gatti2015}, not relying on the Gaussian approximation  (\ref{psi_gauss}), which predicts
a nearly separable two-photon state for
\beq
\taus\ll\tau_p\ll\taui
\eeq

{\bf (ii) Co-propagating geometry with $\eta=0$}\\
 In the co-propagating configuration , a  method to achieve a nearly separable state consists in matching the group velocities
of the signal and the pump  \cite{grice2001}. 
If condition $v_{gs}=v_{gp}$ is satisfied,  one has $\taus=0$ ,  $\eta=0$, and Eq.(\ref{Kgauss}) reduces to
\beq
\cappa^{\sppm}=
1+\frac{1}{2\gamgauss}\left(\frac{\tau_p}{\tauib}\right)^2 
\xrightarrow{\tau_p\ll|\tauib|} 1
%\;\;\;\;\;\text{($v_{gs}=v_{gi}$ case)}
\label{Keta0}
\eeq
Thus a nearly separable state can be achieved only asymptotically , for  a pump duration vanishing small, or in practice much smaller than the GVM time between the idler and the pump, which clearly requires subpicosecond pump pulses. 
Notice that the $\eta=0$ condition can be satisfied also 
in the counter-propagating configuration, where separability is achieved
for much longer pulses satisfying the less stringent requirement $\tau_p\ll\taui$ \cite{gatti2015}.
%%%%%%%%%%%%%%%%%%%%%%%
\begin{table*}
\begin{tabular}{|l|c|c|c|c|c|c|c|c|c|}
 \hline\hline
\textbf{crystal}  &$l_c$ (mm) & phase matching ($\theta_p$)  & $\lambda_p$ &$\lambda_s\,$&$\lambda_i\,$&$\taus$\,  
                    & $\tauipm$ & $\tau_p^{min}$ & $\eta$\\ \hline
{\bf (i)} PPKTP  & 10mm& type 0  e-ee ($90^\circ$) &821.4nm&1141nm& 2932nm  & 0.67ps & 63ps &4.05ps &0.01\\
{\bf (ii)} KDP  & 10mm& type II e-oe ($67.8^\circ$) & 415nm &830nm& 830nm  &  0     & 0.72ps & 0     & 0\\
{\bf (iii)} BBO & 10mm& type II e-oe ($28.8^\circ$) & 757nm &1514nm& 1514nm &  -0.237ps & 0.237ps & 0.147ps &-1 \\
\hline\hline
\end{tabular}
\caption{Phase-matching conditions and characteristic time constants  for the three crystals taken as examples: 
(i) periodically poled KTP, with $800$nm poling period for the counter-propagating configuration, 
(ii)  KDP and (iii)  BBO bulk crystal for the two co-propagating configurations.}\label{table1}
\end{table*}

{\bf (iii) Co-propagating geometry with $\eta=-1$}\\
The symmetric condition $\eta=-1$  can be fulfilled only in the co-propagating configuration,  and is rather difficult to meet because
it requires that the pump group inverse group velocity falls exactly midway between the signal and the idler inverse 
group velocities since
\beq
\tauib=-\taus\longleftrightarrow
\frac{1}{2}\left(\frac{1}{v_{gs}}+\frac{1}{v_{gi}}\right)=\frac{1}{v_{gp}}
\label{sym}
\eeq
Provided this relation is satisfied, the two-photon correlation $\psi(\Omega_s,\Omega_i)$ displays a circular shape 
for $\tau_p=\tau_p^{min}$, since $c_{12}=0$ and $c_{11}=c_{22}=2\gamma\taus^2$.
For the optimized pump pulse duration, the generated twin photons are thus not only uncorrelated but also indistinguishable.
%, a relevant feature required in a number of quantum information protocol based on 
%&Hong-Ou-Mandel type interferometer schemes ????. 
Conditions {\bf (ii)} and {\bf (iii)} are referred to as asymmetric and symmetric group-velocity matching respectively.
They are usually difficult to satisfy in the visible range where normal dispersion implies that the group velocities 
increase with the wavelength. On the other hand, some $\chi^{(2)}$ materials
offers the possibility to achieve group-velocity matching in the near infrared and at telecom wavelengths, 
as shown  in \cite{grice2001,uren2006,migdall2013}.
Experimental evidence of the degeneration of frequency decorrelated photon pairs through this technique
are reported in \cite{mosley2008b}.
\par
Table \ref{table1} summarizes the parameters for three specific examples, chosen as 
representative of  the configurations {\bf (i)}, {\bf (ii)} and {{\bf (iii)}. \\
For the counter-propagating  geometry {\bf (i)}  we consider  a $10$mm long periodically poled crystal  of Potassium Titanyl Phosphate (PPKTP) in a type $0$ (e-ee) phase-matching configuration: the poling period is $\Lambda=800$nm, 
$\lambda_p=814.5$nm, $\lambda_s=1145$nm, $\lambda_i=2932.4$nm, $\eta=\taus/\taui=0.01$
(apart from the length of the crystal,
the parameters are taken from the experiment reported in \cite{canalias2007}).\\
For the co-propagating geometry,  we consider two different bulk negative uniaxial crystals of $10$mm length, both
tuned  to generate a separable state along the collinear direction under appropriate conditions: {\bf (ii)} a   Potassium Dihydrogen Phosphate (KDP) crystal and {\bf (iii)} a Beta-Barium Triborate (BBO) crystal
both cut for type II collinear phase-matching (e-oe) at degeneracy.
When pumped at $415$nm with a tuning angle $\theta_p=68.7^\circ$ with the crystal axis, the KDP crystal
has the peculiarity of displaying
a vanishing GVM between pump and the signal field (i.e. $\taus=0$, $\eta=0$) and is therefore well
suited for the generation a separable two-photon state provided that 
$\tau_p\ll\tauib=0.72$ps \cite{grice2001,mosley2008}. 
For the BBO crystal pumped at $757$nm with a pump tuning angle $\theta_p=28.8^\circ$ 
we have $\taus=-\taui=0.237ps$, $\eta=-1$.  
%%%%%%%%%%%%%%%%%%%%%%%%%%%%%%%%%%%%%%%%%%%%%%%%%%
\begin{figure*} 
\centering
\includegraphics[scale=0.6]{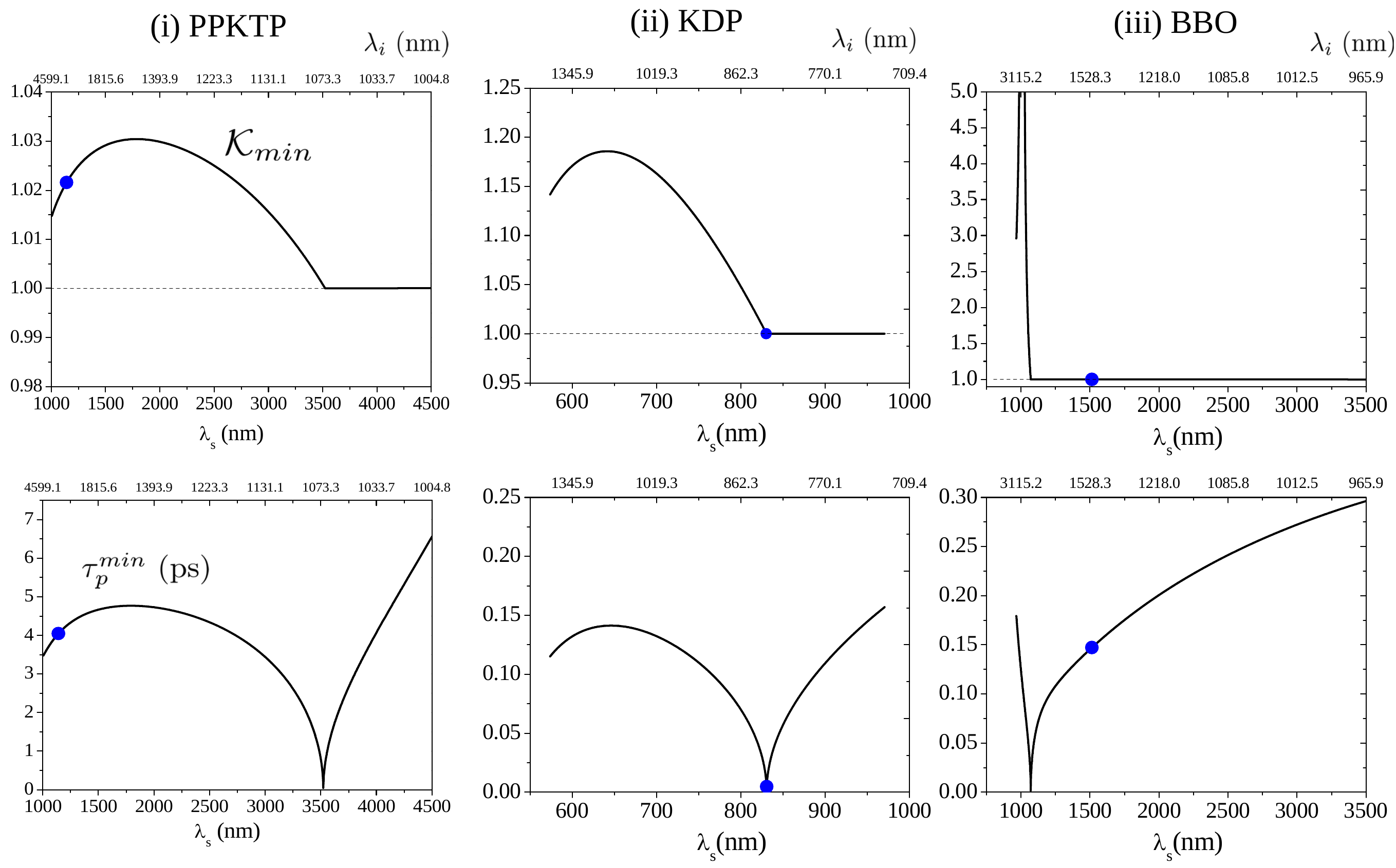}
\caption{Minimum value of the Schmidt number $K_{min}$ (top panels) and relative pump duration $\tau_p^{min}$ (bottom panels) as a function of the signal
wavelengths,  evaluated with the Gaussian approximation [ Eq.(\ref{Kmin}) and (\ref{taupmin})] for the three crystals chosen as examples. The top
horizontal scale shows the conjugate idler wavelength $\lambda_i$.
The  blue dots correspond to the parameters  in Table \ref{table1}}.
\label{figKlambda}
\end{figure*}
\begin{figure} 
\includegraphics[scale=0.75]{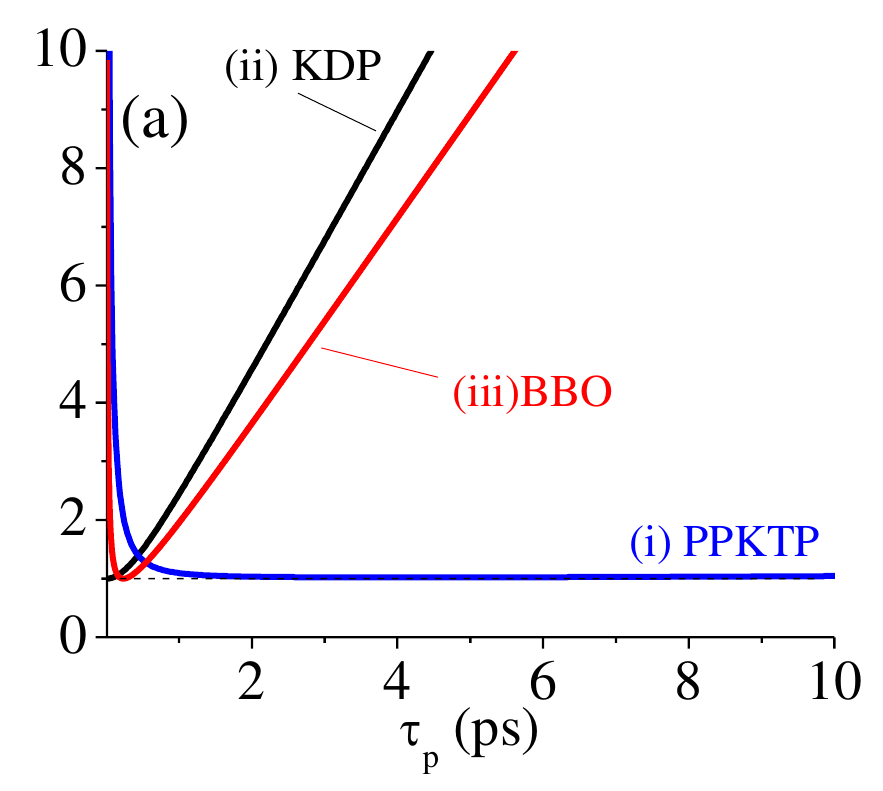}
\caption{Schmidt number $\cappa$ evaluated with the Gaussian approximation (\ref{Kgauss}) 
as a function of the pump pulse duration $\tau_p$ for the three examples in Table.\ref{table1}}
\label{figKtaup}
\end{figure}
\begin{figure*}  
\centering
\includegraphics[scale=0.6]{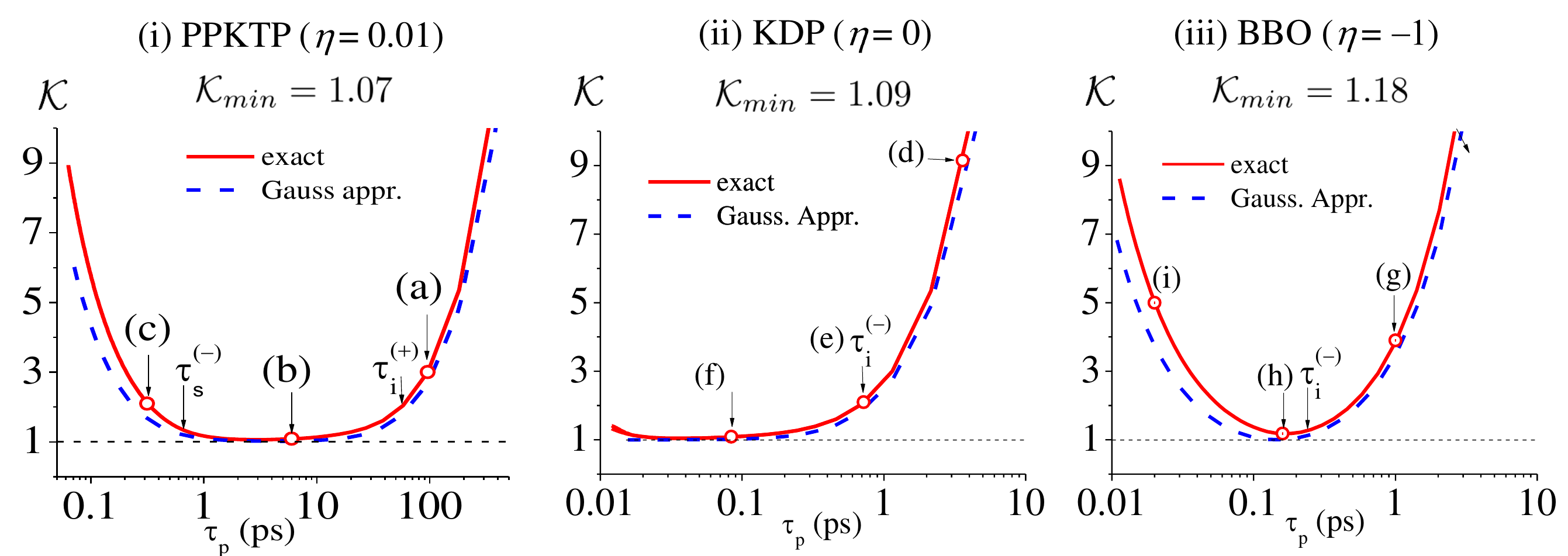}
\caption{Schmidt number $\cappa$ 
as a function of the pump pulse duration $\tau_p$ for the (i)  KTP counter-propagating case,   (ii) KDP co-propagating  and  (iii) BBO 
co-propagating  cases (parameters  in Table \ref{table1}). 
In (i) the state state is nearly separable for pump durations  
$\tau_p$ intermediate between $\taus=0.67$ps and $\taui=63$ps. 
For the co-propagating cases (ii) and (iii), 
separability is achieved only for subpicosecond pulses with $\tau_p\ll\tauib$.  
The minima  of $\cappa$, i.e. the amount of achievable separability, are comparable in the three cases. 
The hollow red dots correspond to the plots shown  in Fig.\ref{fig4}.
}
\label{fig3}
\end{figure*}
\begin{figure*} 
\centering
\includegraphics[scale=0.7]{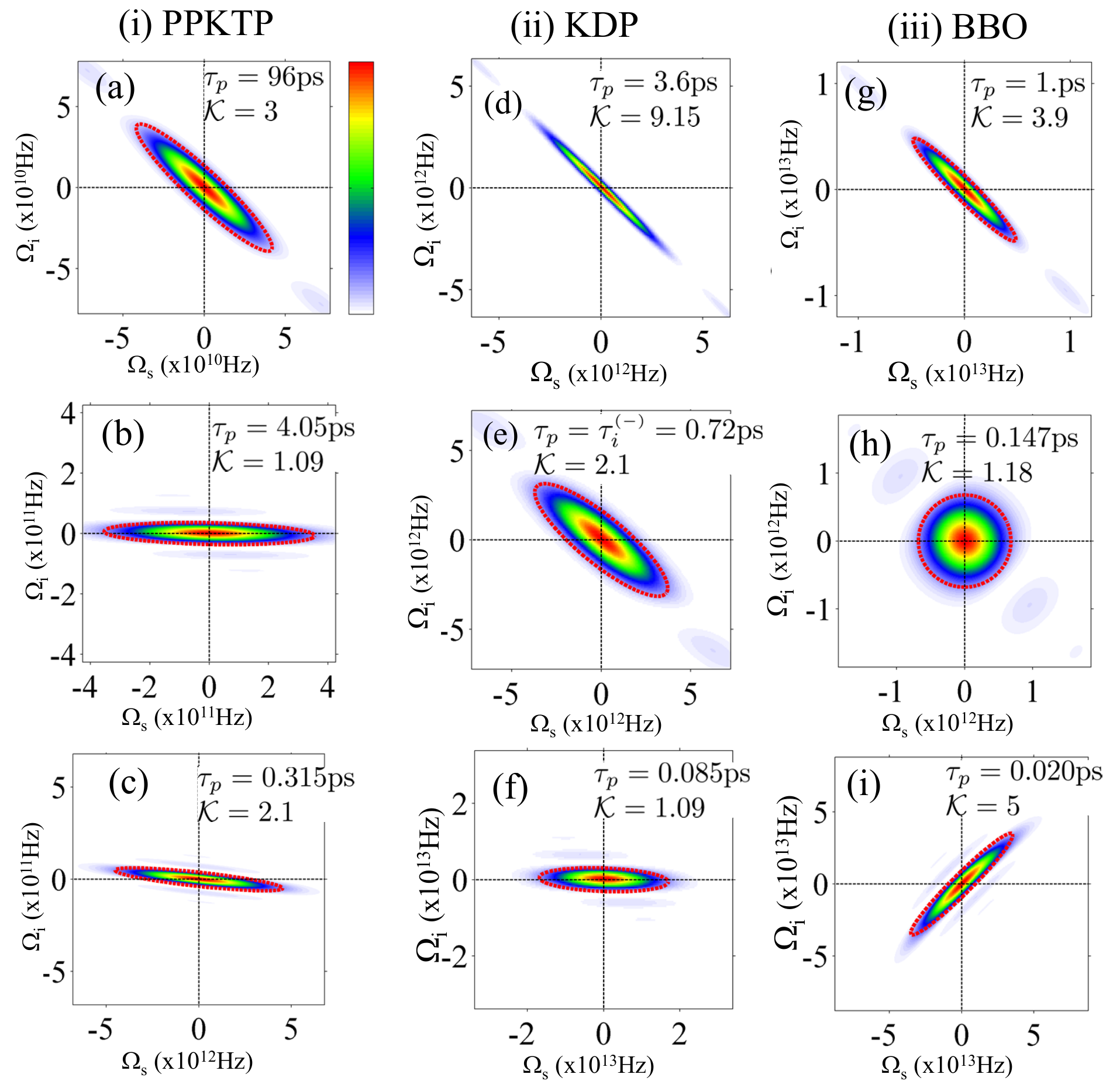}
\caption{Spectral biphoton correlation $|\psi(\Om_s,\Om_i)|^2$ plotted for decreasing pump pulse duration (from top to bottom) corresponding to the hollow red dots shown in Fig.\ref{fig3}. In the short pulse PPKTP case, (c) the idler frequency scale $\Omega_i$ is zoomed by factor 10 with respect to the signal frequency
scale. The red ellipses corresponds to Eq.(\ref{quadr}).}
\label{fig4}
\end{figure*}
\par
Fig. \ref{figKlambda} plots  the results  of the Gaussian approximation for $K_{min}$ and $\tau_p^{min}$ [Eqs.  (\ref{Kmin}) and (\ref{taupmin})], as 
a function of the signal central wavelengths $\lambda_s$, for these three examples. The phase-matched wavelengths, $\lambda_s$ and  $\lambda_i$,
and the corresponding characteristic times $\taus$ and $\tauipm$ are evaluated using the Sellmeier dispersion formula
reported in \cite{niko91,kato2002,zernike64}.
% It should be stressed that only collinear SPDC radiation is taken into account. 
For the PPKTP crystal, 
different wavelengths corresponds to different poling periods $\Lambda$, not reported in the figure.
For the bulk KDP and BBO crystals the signal and idler central wavelengths are varied by changing the tuning angle 
$\theta_p$ between the pump direction and the crystal axis (not reported in the figure). Notice that in
the BBO case $\eta$ is always negative for $\lambda_s>1070\,$nm, so that the generated two-photon state is separable 
when the crystal is tuned on those wavelengths according to approximation (\ref{Kmin}). 
Notice also that
for $\lambda_s=1010\,$nm the group velocities of the signal and idler fields becomes equal ($\eta=1$)
and the Schmidt number predicted by Eq.(\ref{Kgauss}) goes to infinity. 
Under these conditions the SPDC bandwidths and the number entangled modes are in fact very large, through not infinite,  as they are only limited by group-velocity dispersion, a feature not taken into account in the simplified model based 
the linearized phase-matching function (\ref{Dlin}). 

Figures \ref{figKtaup} and \ref{fig3}   illustrate the behaviour of the Schmidt number $\cappa$  as a function of the
the pump pulse duration. The phase-matching conditions for the three crystals are
those reported in Table \ref{table1} and correspond to the blue dots in Fig.\ref{figKlambda}.
Figure \ref{figKtaup} reports the results of the Gaussian approximation  (\ref{Kgauss}) in linear scale (to allow immediate comparison between the three cases), while Fig. \ref{fig3} compares the approximate results  with the more exact ones  obtained
through numerical integration of  Eqs.(\ref{kintegral})-(\ref{B}). In this latter
case dispersion is fully taken into account, 
the phase-mismatch functions (\ref{mismatch}) being evaluated using the complete Sellmeier
 formula. The logarithmic horizontal scale used in this case evidences the different ranges of pump durations  $\tau_p$ 
which must be used for achieving separability  in the various configurations.

From these plots one can notice that in order to achieve separability the two co-propagating configurations require supbicosecond pulses, whose duration must be close to
$\tau_p^{min}=147$fs in the BBO case, $\taup\ll\tauib=720$fs in the KDP case. In contrast, the counter-propagating geometry displays a negligible
amount of  entanglement over a broad plateau ranging from $\tau_p\sim 2$ps up to $\tau_p\sim 10$ps. 
\par
Figure  \ref{fig3} displays some 
discrepancy between the approximated results \eqref{Kgauss} and the exact one especially for short pump pulses,
where dispersion plays an important role  due to the large bandwidths involved
%{\bf NOTA: a me sembra soprattutto questo il motivo!!} 
and the phase-matching function mainly determines the twin photon correlation. 
In particular, the minimum value of $\cappa$ is always slightly larger than the value predicted by the Gaussian result \eqref{Kmin} and  never reaches unity even in the two examples with $\eta \le 0$. Actually, 
the amount of purity which can be achieved in the counter-propagating geometry is comparable
to that of the two other configurations, 
which require much more stringent phase matching conditions and ultra-short pulses. \\
We also notice that  the  sidelobes of the sinc function (clearly visible e.g. in Fig.\ref{fig4}h) are not taken into account by approximation (\ref{sinc_apr}) and lead  to a slight increase of the  amount of entanglement with respect to the prediction of  relation (\ref{Kgauss}) in all cases. 
Branczyk et al. \cite{branczyk2011} demonstrated the possibility
to eliminate the residual entanglement associated to these  sidelobes by modifying
the periodic poling of the $\chi^{(2)}$ nonlinearity in order to produce
a Gaussian-shaped phase matching function. 

Figure \ref{fig4} shows the biphoton correlation $|\psi(\Om_s,\Om_i)|^2$ in the signal-idler frequency plane. 
For each crystals (i), (ii) and (iii),  the pump pulse duration decreases from top to bottom  
(the value of $\tau_p$ corresponds to the large hollow dots shown in Fig.\ref{fig3}). 
The red ellipses superimposed to the density plots show the curve 
\beq
\label{quadr}
c_{11}\Om_s^2+c_{22}\Om_i^2+2c_{12}\Om_s\Om_i= 1
\eeq
where according to the Gaussian formula (\ref{psi_gauss}) $|\psi|^2$ reduces by a factor $1/e^2$. 
Its principal major axis forms an angle $\theta$ with the $\Om_s$-axis given by 
\beq
\theta
=-\frac{1}{2}\arctan\left(\frac{(\tau_p/\tauipm)^2+2\gamma\eta}{\gamma(1-\eta^2)}\right)
\eeq
We have then the following limiting behaviours
\begin{align} 
&\theta  \to  -\frac{\pi}{4}  \;\;\text{for}\;    
\tau_p\gg|\tauipm | 
\\
&\theta  \to -\arctan\eta \;\;\text{for}\;         
\begin{cases} 
\tau_p\ll |\taus| &\text{if}\;\eta \ne 0 \\
\tau_p\ll|\tauipm|  &\text{if}\;\eta = 0 
\end{cases} 
\end{align}
The first limit, with $\tau_p\gg|\tauipm|$, corresponds to a nearly monochromatic pump pulse
with the two-photon state strongly entangled in frequencies. Accordingly, 
the spectral two-photon amplitude $\psi$ develops along the $\Omega_i=-\Omega_s$ diagonal,  where energy conservation takes place (see  panels (a),(d) and (g) 
in Fig.\ref{fig4}).

%{\bf NOTA Riformula la discussione seguente in accordo con il risultato, in particolare verifica sempra i segni, perche' $\taus$  e $ \tauipm$ possono essere sia %positivi che negativi} 
In the counter-propagating geometry of the PPKTP crystal {\bf (i)}, the state appears separable 
(the ellipse has its  axes aligned with $\Omega_s$ and $\Omega_i $) for  pulses
of several picoseconds such that $\taus\ll\taup\ll\taus$, as shown in Fig.\ref{fig4}b.
Only for very short pump pulses with $\tau_p\ll\taus$, phase-matching determines correlation  
with $\psi$ aligned along  the  line $\Om_i=-\eta\Om_s$ (Fig.\ref{fig4}c). 

In the co-propagating configurations {\bf (ii)} and {\bf (iii)}, separability is only achieved for 
pulses in the subpicosecond range satisfying the condition $\tau_p\ll|\tauib|$. In the KDP case the two-photon state remains separable for $\tau_p\rightarrow 0$ as a consequence of the group-velocity matching of the signal and pump photons ($\eta=0$).
For the BBO crystal with symmetrical group velocity matching ($\eta=-1$), the biphoton correlation
displays a nearly circular symmetry for $\tau_p=\tau_p^{min}=147$fs (Fig.\ref{fig4}h).
In this latter case the frequencies becomes again correlated for ultra-short pulses, 
the correlations developing along the $\Oms=\Omi$ diagonal (Fig.\ref{fig4}i).

\section{Spectral properties of twin photons}
\label{sec:spectrum} 
The Gaussian approximation also allows an immediate  comparison  of   the spectral properties of the twin photons generated in the various configurations, by calculating their first order coherence functions. By inserting the Gaussian formula for the biphoton amplitude inside  Eqs. (\ref{G1s},\ref{G2s}), we get 
\begin{align}
G_s^{(1)}(\Oms,\Oms') &\approx
\frac{g^2\tau_p^2}{\sqrt{8\pi c_{22}}}
e^{-i\taus(\Oms-\Oms')}\nn\\
&\times e^{-\frac{2c_{11}c_{22}-c_{12}^2}{2c_{22}}(\Oms^2+\Oms'^2)
+\frac{c_{12}^2}{c_{22}}\Oms\Oms'}\label{g1sgauss}
\\
G_i^{(1)}(\Omi,\Omi') &\approx
\frac{g^2\tau_p^2}{\sqrt{8\pi c_{11}}}
e^{-i\tauipm(\Omi-\Omi')}\nn\\
&\times e^{-\frac{2c_{11}c_{22}-c_{12}^2}{2c_{11}}(\Omi^2+\Omi'^2)
+\frac{c_{12}^2}{c_{11}}\Omi\Omi'}\label{g1igauss}
\end{align}
Using these formulas, we can estimate the bandwidths $\sigma_j$ of the signal and idler spectra
\beq
{\cal S}_j(\Om_j):=G_j^{(1)}(\Om_j,\Om_j)
\propto e^{-\frac{\Om_j^2}{2\sigma_j^2}}
 \;\; (j=i,s).
\eeq
Expliciting the $c_{ij}$ coefficients given in Eqs.\eqref{c11}-\eqref{c12}, we find
\beqa
\sigma_s %=\sqrt{\frac{c_{22}}{c_{11}c_{22}-c_{12}^2}}
&=&\frac{1}{\sqrt{2}(1-\eta)}\left(\frac{1}{2\gamma\tauipm^2}+\frac{1}{\tau_p^2}\right)^{1/2}\\
\sigma_i %=\sqrt{\frac{c_{11}}{c_{11}c_{22}-c_{12}^2}}
&=&\frac{1}{\sqrt{2}(1-\eta)}\left(\frac{1}{2\gamma\tauipm^2}+\frac{\eta^2}{\tau_p^2}\right)^{1/2}
\eeqa
Considering the particular cases in which the state becomes separable or nearly separable (corresponding
to the examples shown in the panels (b), (f) and (h) of Fig.\ref{fig4}), 
we have 
\beqa
\sigma_s 
%&=&\frac{\sqrt{1+|\eta|}}{1-\eta}\frac{2\sqrt{\ln 2}}{\tau_p^{min}}\;\;   \\
&\approx&
\begin{cases}
\frac{1}{\sqrt{2}\tau_p^{min}}  &\text{case {\bf (i)}}:|\eta|\ll 1,\;\tau_p=\tau_p^{min}\\
\frac{1}{\sqrt{2}\tau_p}        &\text{case {\bf (ii)}}:\eta=0,\;\tau_p\ll|\tauib|\\
\frac{1}{2\tau_p^{min}}  &\text{case {\bf (iii)}}:\eta=-1,\;\tau_p=\tau_p^{min}\label{fwhm1}
\end{cases}\\
\sigma_i 
&\approx&
\begin{cases}
\frac{1}{2\sqrt{\gamma}|\taui|}\;\;&\text{case {\bf (i)}}:|\eta|\ll 1,\;\tau_p=\tau_p^{min}\\
\frac{1}{2\sqrt{\gamma}|\tauib|}    &\text{case {\bf (ii)}}:\eta=0, \tau_p\ll |\tauib| \\
\frac{1}{2\tau_p^{min}}\;\;& \text{case {\bf (iii)}}:\eta=-1,\;\tau_p=\tau_p^{min}\label{fwhm2}
\end{cases}
\eeqa
%%%%%%%%%%%%%%%%%%%%%%%%%%%%%%%%%%%%%%
\begin{figure*} 
\centering
\includegraphics[scale=0.65]{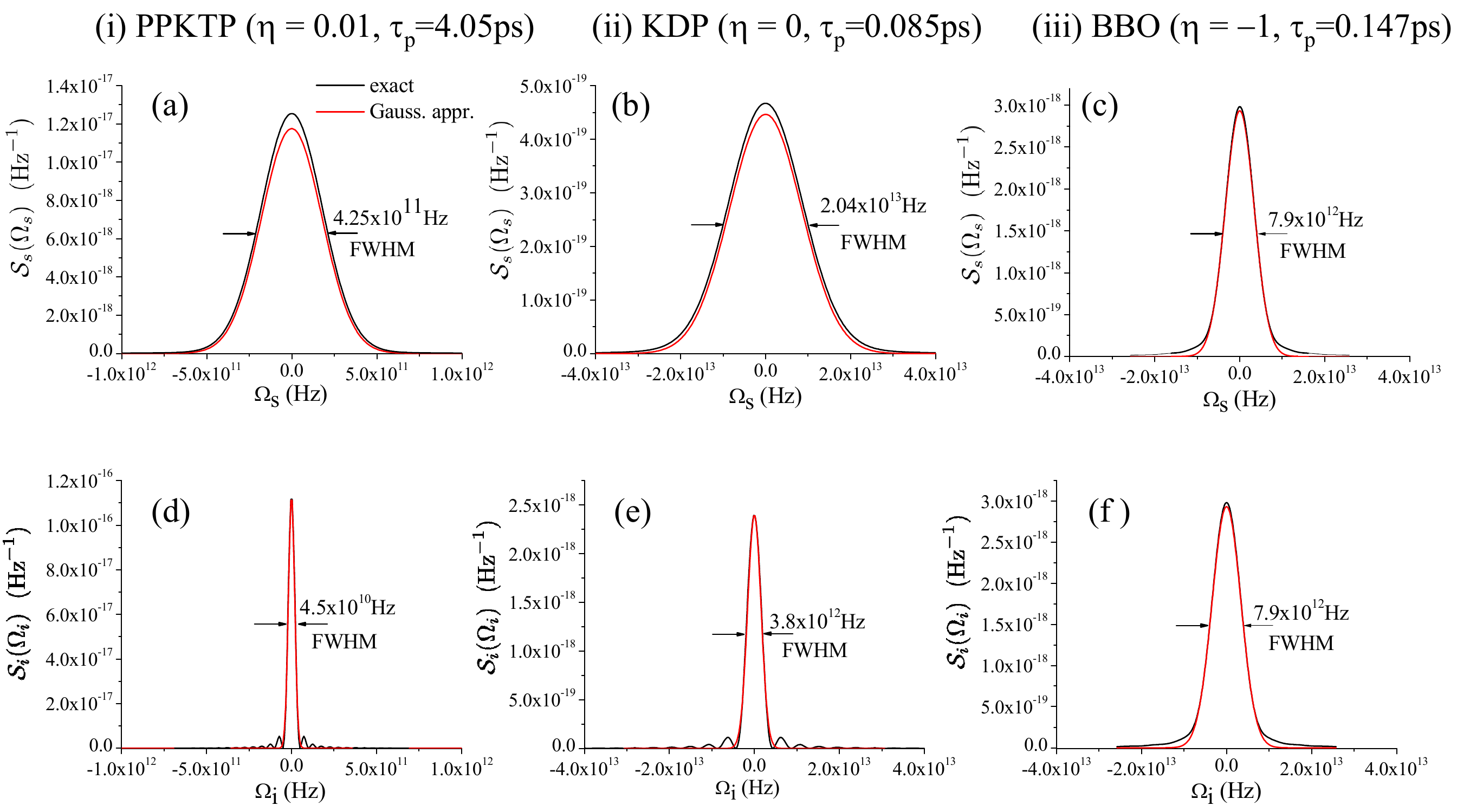}
\caption{Spectra of the signal (top) and of the idler (bottom)  for the three examples considered, in conditions of nearly separability of the state
corresponding to panels (b), (f), and (h) of Fig.\ref{fig4}. Black lines:  numerical results 
 from Eqs.(\ref{G1s}),(\ref{G2s}).  Red lines: Gaussian approximation (\ref{g1sgauss}),(\ref{g1igauss}).  The indicated bandwidths (FWHM)  are calculated  from  the "exact " numerical results and match  the approximated ones  in  Eqs.(\ref{fwhm1})-(\ref{fwhm2})
within an error of less than 10\%. }
\label{fig6}
\end{figure*}
%%%%%%%%%%%%%%%%%%
Fig.\ref{fig6} plots the spectra of the signal and idler photons,  in the optimal conditions for separability, calculated both  with this Gaussian approximation and with the more exact  numerical integration of Eqs. (\ref{G1s})-(\ref{G2s}). From this figure and 
from the approximated results in Eqs. \eqref{fwhm1} \eqref{fwhm2},  we observe that in conditions of separability : 
\begin{enumerate}
\renewcommand{\theenumi}{\roman{enumi}}
\item In all the cases, the bandwidth of the signal photon reproduces basically that of the pump laser  (a part some inessential $\sqrt{2}$ factors). Clearly,  in the counter-propagating configuration it is much narrower (less than ThZ) than in the co-propagating case, because separability is achieved in the former case for longer pump pulses. 
\item The bandwidth  of the idler photon is rather determined by the
phase-matching characteristic time $\tauipm$ (notice that in case ({ iii}),
of symmetric group-velocity matching, the two bandwidths coincides since 
$\tau_p^{min}=\sqrt{2\gamma}|\tauib|$). As expected, the bandwidth of the backward propagating idler 
 is more than two order of magnitude narrower than those of the co-propagating idler photons, because  the GVS characteristic time $\taui$ is two orders of magnitude longer than than the GVM characteristic times involved in the co-propagating cases. 
\end{enumerate}

\section{Conclusions}
In this work we compared different phase-matching configurations suitable for 
generating pure heralded single photons from spontaneous parametric down-conversion. We provided a detailed
analysis of the conditions under which separable twin twin photons can be generated through the quantitative evaluation
of the Schmidt number as a function of the pump pulse duration. 
Because of the natural separation of the GVM and the GVS time  scales $\taus$ and $\taui$, the counter-propagating geometry
offers the advantage of generating separable twin photons without the need to fine tune their relative group-velocities 
as in standard co-propagating configurations.
Because of this unique feature, counter-propagating twin photons in a pure state can in principle be heralded at any frequency 
by choosing the required poling period. Moreover, the twin photons
are naturally narrow band, especially the one propagating opposite to the pump direction, and separability
is achieved for a broad range of pump pulse durations within $\taus$ and $\taui$.

In contrast, twin photons emitted in the common co-propagating geometry are naturally broadband 
and can be generated in a separable state only for very short pulses, under particular phase-matching conditions and at particular wavelengths
depending on the material. The counter-propagating configuration offers thus much more flexibility, once the technical challenges for the fabrication of crystals with sub-micrometric poling periods are overcomed.

%\bibliography{biblio}
%\bibliographystyle{apsrev4-1}

%

\end{document}